\documentclass[12pt]{article}

\newcommand{\be}{\begin{equation}}
\newcommand{\ee}{\end{equation}}
\newcommand{\beq}{\begin{eqnarray}}
\newcommand{\eeq}{\end{eqnarray}}
\newcommand{\sla}[1]{\not \!#1}
\newcommand{\kint}{\int \frac{{\rm d}^4 k}{{\left(2 \pi \right)}^4}\, }
\newcommand{\tr}[2]{{\rm tr}_{#1}\Big[ #2 \Big]}
\newcommand{\hk}[1]{\,{\rm #1}}
\def\Dirac#1{#1\hskip-6pt/}
\newcommand{\partialboth}{\stackrel{\leftrightarrow}{\nabla}}

\begin{document}
\rightline{RUB-TPII-03/00}
\begin{center}
{\Large
Heavy quark mass expansion
and intrinsic charm in light hadrons }\\
\vspace{0.4cm}
{\bf M. Franz$^a$, M.V. Polyakov$^{a,b}$,
K. Goeke$^{a}$}\\
\vspace{0.4cm}
{\it
$^a$Institut f\"ur Theoretische Physik II, Ruhr--Universit\"at Bochum,\\
 D--44780 Bochum, Germany\\
$^b$Petersburg Nuclear Physics Institute, 188350 Gatchina, Russia}
\end{center}
\vspace{0.1cm}

\begin{abstract}
We review the technique of heavy quark mass expansion
of various operators made of heavy quark fields
using a semiclassical approximation.
It corresponds to an operator
product expansion in the form of series in the inverse heavy quark mass.
This technique applied recently to the axial current
is used to estimate the
charm content of the \( \eta  \), \( \eta \prime  \) mesons and the intrinsic
charm contribution to the proton spin.
The derivation of heavy quark mass expansion for \( \langle
\bar{Q}\gamma_{5}Q\rangle \) is given here in detail and the expansions
of the scalar, vector and tensor current and of $\langle
\bar{Q}\nabla_\mu \gamma_\nu Q \rangle$ (a~contribution to the
energy-momentum tensor) are presented as well. The obtained
results are used to estimate the intrinsic charm contribution
to various observables.

\end{abstract}
\vspace{0.3cm}

\section{Introduction}
\label{sec:intod}

Nowdays it is established beyond any doubts that the
naive picture of light hadrons as made of three
constituent quarks (for baryons)
or $q\bar q$ pairs  of constituent quarks (for mesons) is not complete.
The DIS experiments revealed the rich sea structure of the nucleon,
these experiments showed in particular that a considerable portion of the
nucleon spin is carried by the strange component of the nucleon sea.
Furthemore there are experimental facts which seems to suggest that
a non-vanishing nonperturbative component of intrinsic charm is present
in light hadrons \cite{C1,C2}.

We address the problem of intrinsic charm content of light hadrons from
the point of view of the heavy quark mass expansion.
The $c\bar c$ pairs in light
hadrons, due to parametrically large mass of charm quarks,
can appear in a light hadron as virtual state whose life time is short,
of order $1/m_c$. The nonperturbative (with typical momenta
below heavy quark mass $m_c$)
gluon and light quark fluctuations are slowly varying from ``point view"
of virtual $c\bar c$ pair, hence the heavy quark mass expansion is
equivalent to the semiclassical expansion. This expansion allows
to rewrite operators made of heavy quarks in terms of light degrees
of freedom (gluons and light quarks). For a detailed
discussion of the heavy quark mass expansion see ref.~\cite{Novikov}.

Let us note also that in absence of a direct probe of gluons the open
charm production is considered as the main source of information about
nucleon's gluon distributions.
In hard leptoproduction
heavy quarks are produced in the
leading order via the photon-gluon
fusion (PGF). The leading graph for PGF can be related directly
to gluon distributions
 {\it if one assumes} that there is no intrinsic
charm content in the nucleon (no $c(x), \bar c(x)$ and
no $\Delta c(x), \Delta \bar c(x)$ at normalzation point $\mu=m_c$).
However now there are many evidences that, in principle, there
might be considerable intrinsic charm component in the nucleon wave
function even at low normalization point.
For reliable extraction of gluon distributions from open charm
electroproduction experiments it is necessary to have quantitative
estimates of the intrinsic charm content of the nucleon.

This paper will be organized as follows:
In the first part we present
the calculation of the expectation value of
heavy quark currents in the background of gluon fields using
a semiclassical approximation.
 This corresponds to an expansion in the inverse of the heavy
quark mass
\be
\langle Q^\dagger(x) \Gamma Q(x) \rangle
= \sum_n {1 \over m^n} X_\Gamma^{(n)} \, ,
\ee
where $\Gamma$ denotes the Lorentz structure of the
current and the $X_\Gamma^{(n)}$ are
local expressions of the field strength depending
on $\Gamma$.
In section \ref{secdet} we review the large m expansion of
the fermion determinant
appearing in our definition of the expectation value.
In section \ref{genexp} we then outline the expansion of color singlet
currents in general before we present in section \ref{axiexp} the
expansion of the axial current used in \cite{FraPobPolGoe99} in full detail.
The connection to the expectation value of the axial vector current using
the axial anomaly equation and some general restriction coming from this
equation are given in section~\ref{axivecexp}.
As further examples we present the expansion of the scalar current
in section \ref{scaexp}, the vector current (section \ref{vecexp})
and the tensor current (section~\ref{tenexp}), respectively.
In section \ref{enemomexp} we finally show the result of the expansion
of $\langle Q^\dagger(x) \nabla_\mu \gamma_\nu Q(x) \rangle$, appearing
in the energy-momentum tensor of QCD.

In the second part we discuss the calculation of intrinsic heavy
quark content of light hadrons as an application of the heavy quark
mass expansion. In the case of charm content of $\eta',\eta$ mesons
and intrinsic charm contributions to the proton spin we reduce the
calculations of these quantities to matrix elements which are already
known either phenomenologically or were computed previously.
In other cases, like intrinsic charm contribution to the nucleon
tensor charge and to energy momentum tensor, the problem is reduced
to matrix elements of gluon operators which can be estimated
using various nonperturbative methods in QCD:
lattice calculation, QCD sum rule, theory of instanton vacuum.

\section{Heavy quark expansion of
currents in the background gluon and light quark fields}
The expectation value of a color-singlet quark
current made of heavy quarks
in the background of gluon and light quark fields
can be written after integration out heavy degrees of freedom as:
\beq
\langle Q^\dagger(x) \Gamma Q(x) \rangle &=&
\hk{det} D \,
\hk{tr}_{c,\gamma}\langle x | {1 \over D }
\Gamma | x \rangle \;,
\label{eq:det*tr}
\eeq
Here $\Gamma$ denotes an arbitrary Lorentz-structure.
Note that all calculations will be performed in the euclidean
space-time, so the QCD Dirac operator reads:
\begin{equation}
D=\hk{i}\! \sla{\nabla} + \hk{i} m \,
\end{equation}
where the covariant derivative is defined as
\be
\nabla_\mu = \left( \partial_\mu - \hk{i} {\lambda^a \over 2} A^a_\mu(x) \right)
\ee
and $m$ is the heavy quark mass.
For the used conventions and the euclidization see the appendix.
Eq.(\ref{eq:det*tr}) can now be expanded in a power-series of
the inverse heavy quark mass, $1/m$,
under the assumption that the gradient of the
background field strength is small compared
to $m$.
The expansion of determinant of the Dirac operator in eq.(\ref{eq:det*tr})
has been calculated
by a large number of authors, see e.g.
\cite{AndGro78,VaiZakNovShi84,DPY}.
We briefly review the calculation of the determinant following
\cite{DPY} since
we use the result of this calculation as a check of the expansion of
a scalar current of heavy quarks in Section \ref{scaexp}.

\subsection{Expansion of the determinant}
\label{secdet}
The expansion of the determinant for heavy quarks
in eq.(\ref{eq:det*tr}) yields
divergences of various types. Since most of these divergences are
connected with the determinant of the free Dirac operator we normalize
the determinant with that in zero external field.
For the remaining infinity which can be related  to the
logarithmic renormalization of the coupling constant,
we use the so-called $\zeta$-regularization.
Using the identity
\be
\hk{det} D = \hk{det} D^\dagger
= \left( \hk{det} D^\dagger D \right)^\frac12
\ee
the normalized and regularized determinant can be written as follows:
\be
\label{eq:regdet}
{ \left( { \hk{det} D \over \hk{det} D_0 } \right)
}_{\hk{\zeta-reg}}
 =
\hk{exp} \left[ - {1 \over 2} \lim_{s \rightarrow 0}
\, { \hk{d} \over \hk{d}s }
 \, { M^{2s} \over \Gamma (s) } \int_0^\infty \hk{d} t \,
t^{s-1} \, \hk{Tr}
\left[ e^{-t D^\dag D} - e^{-t D_0^\dag D_0} \right] \right]
\; ,
\ee
where $D_0$ denotes the Dirac operator in the absence of external gluon fields
and $M$ is the regulator mass.
The functional trace denoted by Tr in eq.(\ref{eq:regdet}) can be calculated
with respect to any complete set of states.
For further calculations it is convinient to compute
functional traces in the basis of plane waves, so that
\beq \hk{Tr} \Big[ e^{-t D^\dag D} \Big] &=&
\hk{tr}_{c,\gamma} \int \hk{d}^4 x \kint
e^{-\hk{i}kx} \Big[ e^{-t D^\dag D} \Big]
e^{\hk{i}kx}
\nonumber \\
&=&
\hk{tr}_{c,\gamma} \int \hk{d}^4 x \kint
\Big[ e^{-t D^\dag(\partial_\mu \rightarrow \partial_\mu + \hk{i}k_\mu )
 D(\partial_\nu \rightarrow \partial_\nu + \hk{i}k_\nu ) } \Big]
\cdot 1\;.
\label{eq:unity}
\eeq
The unity in eq.(\ref{eq:unity}) points out that the operators here act on
unity, so that $\partial_\mu \cdot 1=0$.
The further calculations are straightforward: the expression
in eq.(\ref{eq:unity}) can be expanded in powers of the covariant derivative,
integrated with respect to $k$ and the Lorentz indices summed.

Since the explicit calculation is given in \cite{DPY} we present here
only some useful formulae
and the final result for the determinant up to
order ${\cal O}(1/m^2)$.
The square of the Dirac operator in eq.(\ref{eq:unity})
with all differentiation operators
shifted, $\partial_\mu \rightarrow \partial_\mu + \hk{i}k_\mu$,
is given by
\be
\label{eq:DD}
D^\dag(\partial_\mu \rightarrow \partial_\mu + \hk{i}k_\mu )
 D(\partial_\nu \rightarrow \partial_\nu + \hk{i}k_\nu )
= - \nabla^2 + {\sigma \over 2} F - 2 \hk{i} k\nabla + k^2 + m^2 \; ,
\ee
where we have used that
\beq
& F^a_{\mu \nu} & = \hk{i} \left[ \nabla_\mu , \nabla_\nu \right]^a
=
\partial _\mu A^a_\nu - \partial_\nu A^a_\mu + f^{abc} A^b_\mu A^c_\nu     \\
\Rightarrow &
- \sla{\nabla}\sla{\nabla} & = - \nabla^2 + {\sigma \over 2} F \; ,
\eeq
with the notations
$\sigma F=\sigma_{\mu \nu} {\lambda^a \over 2} F^a_{\mu \nu}$
and $\sigma_{\mu \nu}=
\frac{\hk{i}}{2} \left[ \gamma_\mu , \gamma_\nu \right]$ applied.
Expanding the exponential function in eq.(\ref{eq:unity}) the
functional trace then reads
\beq
&& \hspace{-16pt} \hk{Tr} \left[ e^{-t D^\dag D} \right]
= \hk{tr}_{c,\gamma} \int \hk{d}^4 x \kint e^{-t(k^2+m^2)}
\sum_{n=0}^{\infty} (-1)^n { t^n \over n! }
{\left( - \nabla^2 + {\sigma \over 2} F - 2 \hk{i} k\nabla \right)}^n
\! \cdot 1 \hspace{20pt}  \\
&& \hspace{-16pt} = {1 \over 4 \pi^2}
\hk{tr}_{c} \int \hk{d}^4 x\, e^{-tm^2} \left[ \, \frac{1}{t^2}
+ t^0  \left( \frac{1}{6}
\nabla_\alpha \nabla_\beta \nabla_\alpha \nabla_\beta
- \frac{1}{6} \nabla_\alpha \nabla^2 \nabla_\alpha
+ \frac{1}{4} F_{\alpha \beta} F_{\alpha \beta} \right) \right. \nonumber \\
&& + \,t \left( {1 \over 180}
 \nabla^2 \nabla^2 \nabla^2
- {1 \over 36}
 \left( \nabla_\alpha \nabla^2 \nabla_\alpha \nabla^2
+ \nabla_\alpha \nabla^2 \nabla^2 \nabla_\alpha
+ \nabla^2 \nabla_\alpha \nabla^2 \nabla_\alpha \right) \right.
\nonumber \\
&& \hspace{20pt}+ {1 \over 45} \left(
\nabla_\alpha \nabla_\beta \nabla_\alpha \nabla_\beta \nabla^2
+\nabla_\alpha \nabla_\beta \nabla_\alpha \nabla^2 \nabla_\beta
+\nabla_\alpha \nabla_\beta \nabla^2 \nabla_\alpha \nabla_\beta \right.
\nonumber \\
&& \hspace{20pt} \hspace{18pt}\left.
+\nabla_\alpha \nabla^2 \nabla_\beta \nabla_\alpha \nabla_\beta
+\nabla^2 \nabla_\alpha \nabla_\beta \nabla_\alpha \nabla_\beta
+\nabla_\alpha \nabla_\beta \nabla^2 \nabla_\beta \nabla_\alpha \right)
\nonumber \\
&& \hspace{20pt}- {1 \over 90} \left(
\nabla_\alpha \nabla_\beta \nabla_\alpha
\nabla_\gamma \nabla_\beta  \nabla_\gamma
+\nabla_\alpha \nabla_\beta  \nabla_\gamma
\nabla_\alpha \nabla_\beta  \nabla_\gamma \right.
\nonumber \\
&& \hspace{20pt} \hspace{18pt}\left.
+\nabla_\alpha \nabla_\beta  \nabla_\gamma
\nabla_\alpha \nabla_\gamma  \nabla_\beta
+\nabla_\alpha \nabla_\beta  \nabla_\gamma
\nabla_\beta \nabla_\alpha  \nabla_\gamma
+\nabla_\alpha \nabla_\beta  \nabla_\gamma
\nabla_\beta \nabla_\gamma  \nabla_\alpha \right)
\nonumber \\
&& \hspace{20pt}+ {1 \over 6} \left( F_{\alpha \beta} F_{\alpha \beta} \nabla^2
+ F_{\alpha \beta} \nabla^2 F_{\alpha \beta}
+ \nabla^2 F_{\alpha \beta} F_{\alpha \beta}  \right.
\nonumber \\
&& \hspace{20pt} \hspace{18pt} \left.
- \nabla_\gamma F_{\alpha \beta} \nabla_\gamma F_{\alpha \beta}
- \nabla_\gamma F_{\alpha \beta} F_{\alpha \beta} \nabla_\gamma
-F_{\alpha \beta} \nabla_\gamma F_{\alpha \beta} \nabla_\gamma \right)
\nonumber \\
&& \hspace{20pt} \left. \left. - { \hk{i} \over 6}
F_{\alpha \beta} F_{\gamma \beta} F_{\gamma \alpha} \right)
+ {\cal O}(\nabla^8) \right]
\;.
\eeq
Here we used that all
contributions with an odd number of $k$'s vanish
whereas all other integrals
with respect to $k$ yield
\be
\kint k_{\mu_1} \dots k_{\mu_{2n}} e^{-t(k^2+m^2)}
= {1 \over 4 \pi^2} \left( 2 t \right)^{-(n+2)} e^{-tm^2}
\delta_{\mu_1 \dots \mu_{2n}}\; ,
\ee
with $\delta_{\mu_1 \dots \mu_{2n}}$ denoting all possible contractions:
\beq
\delta_{\mu_1 \dots \mu_{2n}}=\exp\biggl[ \frac 12\
\frac{\partial^2}{
\partial \phi_\nu \partial \phi_\nu}\biggr]\
\phi_{\mu_1} \dots \phi_{\mu_{2n}}\Bigr|_{\phi=0}  .
\eeq
After rearranging the terms into gauge invariants and taking also the part
without external gluon fields in eq.(\ref{eq:regdet}) into account,
the determinant up to order
$1/m^2$ can be written as follows:
\beq
{ \left( { \hk{det} D\over \hk{det} D_0} \right)
}_{\hk{\zeta-reg}}
& = &
\hk{exp}\left[ \int \hk{d}^4 x
\left( - {1 \over 48 \pi^2 } \ln \left( {M^2 \over m^2} \right)
\hk{tr}_c F_{\alpha \beta} F_{\alpha \beta} \right. \right.
\nonumber \\
&& \hspace{-64pt}
- { \hk{i} \over 720 \pi^2 } {1 \over m^2} \hk{tr}_c
F_{\alpha \beta} F_{\beta \gamma} F_{\gamma \alpha}
+ { 1 \over 1440 \pi^2 }
{1 \over m^2} \hk{tr}_c
\Big[ \nabla_{\alpha} , F_{\alpha \beta} \Big]
\Big[ \nabla_{\gamma} , F_{\gamma \beta} \Big]
\nonumber \\
&& \hspace{-64pt}
- { 11 \over 1440 \pi^2 }
{1 \over m^2} \hk{tr}_c
\Big[ \nabla_{\gamma} ,
\Big[ \nabla_{\alpha} , F_{\alpha \beta} \Big] \Big] F_{\gamma \beta}
+ { 1 \over 360 \pi^2 }
{1 \over m^2} \partial_\alpha \hk{tr}_c
\Big[ \nabla_{\alpha} , F_{\gamma \beta} \Big]
F_{\gamma \beta}
\nonumber \\
&& \hspace{-64pt} \left. \left.
- { 1 \over  384 \pi^2 }
{1 \over m^2} \partial^2 \hk{tr}_c
F_{\gamma \beta}
F_{\gamma \beta} \right)
+ {\cal O}\left( {1 \over m^4} \right) \right]
\label{eq:detall} \\ && \hspace{-80pt} = \;
\hk{exp}\left[ \int \hk{d}^4 x
\left( - {1 \over 48 \pi^2 } \ln \left( {M^2 \over m^2} \right)
\hk{tr}_c F_{\alpha \beta} F_{\alpha \beta} \right. \right.
\nonumber \\
&& \hspace{-64pt}
- { \hk{i} \over 720 \pi^2 } {1 \over m^2} \hk{tr}_c
F_{\alpha \beta} F_{\beta \gamma} F_{\gamma \alpha}
\nonumber \\
&& \hspace{-64pt} \left. \left. + { 1 \over 120 \pi^2 }
{1 \over m^2} \hk{tr}_c
\Big[ \nabla_{\alpha} , F_{\alpha \beta} \Big]
\Big[ \nabla_{\gamma} , F_{\gamma \beta} \Big]
\right)
+ {\cal O}\left( {1 \over m^4} \right) \right] \;.
\label{eq:detres}
\eeq
Note that in the last step partial integration has been used with all total
derivatives left out.
The effective action $S_{\rm eff,E}=- \ln {\rm det} D$ which can be yielded
from (\ref{eq:detres}),
rotated back to Minkowski space, corresponds exactly to the
result of \cite{VaiZakNovShi84,DPY}:
\beq
S_{\rm eff,M}&=&- {1 \over 48 \pi^2 } \int \hk{d}^4 x
\Big( \ln \left( {M^2 \over m^2} \right)
\hk{tr}_c F_{\alpha \beta} F^{\alpha \beta}
\nonumber \\ && \hspace{88pt}
- { \hk{i} \over 15 \pi^2 } {1 \over m^2} \hk{tr}_c
F_{\alpha \beta} {F^\beta}_{\gamma} F^{\gamma \alpha} \Big)
+ {\cal O}\left( {1 \over m^4}\right) \, ,
\eeq
where equation of motion terms, which vanish in pure Yang-Mills theory
$\Big[ \nabla_{\alpha} , F_{\alpha \beta} \Big] = 0$ have been neglected.

\subsection{Expansion of heavy quark currents}
\label{genexp}
In order to expand $\hk{tr}_{c,\gamma}\langle x | {1 \over D }
\Gamma | x \rangle$ in eq.(\ref{eq:det*tr}) in a series of the inverse heavy
quark mass we can use eq.(\ref{eq:DD}) to  rewrite it as:
\beq
\hk{tr}_{c,\gamma}\langle x | {1 \over D }
\Gamma | x \rangle
&=& \hk{tr}_{c,\gamma} \kint  e^{-\hk{i}kx} {1 \over D^\dag D} D^\dag
\Gamma e^{\hk{i}kx}
\nonumber \\
&& \hspace{-48pt} = \;
\hk{tr}_{c,\gamma} \kint {1 \over k^2 + m^2}
\sum_{n=0}^{\infty} {\left( { \nabla^2 - {\sigma \over 2} F + 2 \hk{i} k\nabla
\over k^2 + m^2} \right)}^n
\left( \hk{i}\! \sla{\nabla} - \sla{k} - \hk{i} m \right) \Gamma
\cdot 1 \;. \hspace{16pt}
\label{eq:fullseries}
\eeq
The expansion in eq.(\ref{eq:fullseries}) is again justified for
small gradients of the gluonic fields
compared to the heavy quark mass $m$.
Depending on the Lorentz structure $\Gamma$ some of the integrals
might be divergent and need to be regularized,
we choose the dimensional regularization,
since the integrals in eq.(\ref{eq:fullseries}) can then be calculated
using:
\be
\label{eq:kint}
\int \frac{{\rm d}^d k}{{\left(2 \pi \right)}^d}\,
{k_{\mu_1} k_{\mu_2} \dots k_{\mu_{2m}} \over
{\left( k^2 + m^2 \right)}^n }
= {1 \over {\left( 4 \pi \right)}^{d/2} }
 { \Gamma\left(n-m-d/2\right) \over \Gamma\left( n \right) 2^m }
\delta_{\mu_1 \dots \mu_{2m}} {\left( {1 \over m^2} \right)}^{n-m-d/2}
\; .
\ee
The number of terms contributing to a given order of $1/m$
is reduced by the fact that terms containing an odd number of
$\gamma$ matrices or an odd number of $k$'s vanish due to the
trace over Lorentz-indices and the integration with respect to $k$.
The expansion of eq.(\ref{eq:fullseries}) then is straightforward.
The result of the expansion must be gauge invariant because we
expand the gauge invariant operator. In order to obtain explicitely
gauge invariant result for heavy quark mass expansion
a number of helpful identities based
on the Bianchi identity:
\be
\label{eq:Bianchi}
\Big[ \nabla_{\alpha}, F_{\beta \gamma} \Big]
+ \Big[ \nabla_\gamma, F_{\alpha \beta} \Big]
+ \Big[ \nabla_\beta, F_{\gamma \alpha} \Big]  = 0 \;,
\ee
can be derived. They will be presented in the following sections.

We want to illustrate some technical details of expanding heavy quark
currents on the example of the pseudoscalar
density and the divergency of the
axial-vector current, which are related to each other
by the axial anomaly. Another motivation to show detailed calculation
for these cases is that recently confusing results for these
cases were reported in the literature
\cite{HalZhi97,AraMusTok98,AraMusTok98b}.
Further we present the result of the
expansion of scalar, vector and tensor currents and of $\langle \bar{Q}
\nabla_\mu \gamma_\nu Q \rangle$ appearing in the energy-momentum
tensor of QCD.

\subsubsection{The pseudoscalar density}
\label{axiexp}
For $\Gamma=\gamma_5$ the expansion eq.(\ref{eq:fullseries})
has the form:
\be
\hk{tr}_{c,\gamma}\langle x | {1 \over D }
\gamma_5 | x \rangle
 = - \hk{i} m
\hk{tr}_{c,\gamma} \kint {1 \over k^2 + m^2}
\sum_{n=0}^{\infty} {\left( { \nabla^2 - {\sigma \over 2} F + 2 \hk{i} k\nabla
\over k^2 + m^2} \right)}^n
 \gamma_5
\cdot 1 \;
\label{eq:fullseriesG5}
\ee
Collecting all terms which contribute up to
${\cal O}(1/m^3)$ one gets:
\beq
\hk{tr}_{c,\gamma}\langle x | {1 \over D }
\gamma_5 | x \rangle
&=& - \hk{i} m \hk{tr}_{c,\gamma} \kint \Big[
{1 \over {\left( k^2 + m^2 \right)}^3} {1 \over 4} \sigma F \sigma F \gamma_5
\nonumber \\
&& \hspace{16pt} \hspace{-80pt}
+ {1 \over {\left( k^2 + m^2 \right)}^4}
\left(
 {1 \over 4} \nabla^2 \sigma F \sigma F
+ {1 \over 4} \sigma F \nabla^2 \sigma F
+ {1 \over 4} \sigma F \sigma F \nabla^2
- {1 \over 8} \sigma F \sigma F \sigma F \right) \gamma_5
\nonumber \\
&& \hspace{16pt} \hspace{-80pt}
- {1 \over {\left( k^2 + m^2 \right)}^5}
\Big(
 \sigma F \sigma F k \nabla k \nabla
+ \sigma F k \nabla \sigma F k \nabla
+ \sigma F k \nabla k \nabla \sigma F
\nonumber \\
&& \hspace{-16pt}
+ k \nabla \sigma F k \nabla \sigma F
+ k \nabla k \nabla \sigma F \sigma F
+ k \nabla \sigma F \sigma F k \nabla \Big) \gamma_5 \Big]
+ {\cal O}\left({1 \over m^5}\right)
\\
&& \hspace{-80pt} = \;
{ \hk{i} \over 32 \pi^2 m } \varepsilon_{\alpha \beta \gamma \delta}
\hk{tr}_c F_{\alpha \beta} F_{\gamma \delta}
- { 1 \over 48 \pi^2 m^3 }
\varepsilon_{\alpha \beta \gamma \delta}
\hk{tr}_c  F_{\rho \alpha} F_{\rho \beta} F_{\gamma \delta}
\nonumber \\
&& \hspace{-64pt}
+ { \hk{i} \over 192 \pi^2 m^3}
\varepsilon_{\alpha \beta \gamma \delta}
\hk{tr}_c \Big[
  F_{\alpha \beta} F_{\gamma \delta} \nabla^2
- F_{\alpha \beta} \nabla_\rho F_{\gamma \delta} \nabla_\rho
+ F_{\alpha \beta} \nabla^2 F_{\gamma \delta}
\nonumber \\ && \hspace{48pt} \hspace{-64pt}
- \nabla_\rho F_{\alpha \beta} \nabla_\rho F_{\gamma \delta}
+ \nabla^2 F_{\alpha \beta} F_{\gamma \delta}
- \nabla_\rho F_{\alpha \beta} F_{\gamma \delta} \nabla_\rho \Big]
+ {\cal O}\left({1 \over m^5}\right)  \; .
\label{eq:intermedres1}
\eeq
Here we have used eq.(\ref{eq:kint}) for the integration over $k$
and the following results for Dirac traces:
\beq
F_{\alpha \beta} F_{\gamma \delta} \tr{\gamma}{\sigma_{\alpha \beta}
\sigma_{\gamma \delta} \gamma_5}
&=& - F_{\alpha \beta} F_{\gamma \delta}
\tr{\gamma}{\gamma_\alpha \gamma_\beta \gamma_\gamma
\gamma_\delta \gamma_5}
\nonumber \\
&=& - 4 \varepsilon_{\alpha \beta \gamma \delta}
F_{\alpha \beta} F_{\gamma \delta} \, ,
\\
F_{\alpha \beta} F_{\gamma \delta} F_{\epsilon \varphi}
\tr{\gamma}{\sigma_{\alpha \beta}
\sigma_{\gamma \delta} \sigma_{\epsilon \varphi} \gamma_5}
&=&
- \hk{i} F_{\alpha \beta} F_{\gamma \delta} F_{\epsilon \varphi}
\tr{\gamma}{\gamma_\alpha \gamma_\beta \gamma_\gamma
\gamma_\delta \gamma_\epsilon \gamma_\varphi \gamma_5}
\nonumber \\
&=& 16 \hk{i} \varepsilon_{\alpha \beta \gamma \delta}
F_{\rho \alpha} F_{\rho \beta} F_{\gamma \delta}
\eeq
Using the following identities:
\beq
\Big[ \nabla_\rho ,  F_{\alpha \beta} \Big]
\Big[ \nabla_\rho ,  F_{\gamma \delta} \Big]
&=&
\nabla_\rho F_{\alpha \beta} \nabla_\rho F_{\gamma \delta}
+ F_{\alpha \beta} \nabla_\rho F_{\gamma \delta} \nabla_\rho
- \nabla_\rho F_{\alpha \beta} F_{\gamma \delta} \nabla_\rho
- F_{\alpha \beta} \nabla^2 F_{\gamma \delta} \, , \;\;\;
\\
\Big[ \nabla_\rho , \Big[ \nabla_\rho , F_{\alpha \beta} \Big] \Big]
F_{\gamma \delta} &=&
\nabla^2 F_{\alpha \beta} F_{\gamma \delta}
+F_{\alpha \beta} \nabla^2 F_{\gamma \delta}
-2\nabla_\rho F_{\alpha \beta} \nabla_\rho F_{\gamma \delta} \, ,
\\
F_{\alpha \beta} \Big[ \nabla_\rho , \Big[ \nabla_\rho ,
F_{\gamma \delta} \Big] \Big] &=&
F_{\alpha \beta} \nabla^2 F_{\gamma \delta}
+ F_{\alpha \beta} F_{\gamma \delta} \nabla^2
-2 F_{\alpha \beta} \nabla_\rho F_{\gamma \delta} \nabla_\rho
\eeq
eq.(\ref{eq:intermedres1}) can be written as
\beq
\hk{tr}_{c,\gamma}\langle x | {1 \over D }
\gamma_5 | x \rangle
&=&
{ \hk{i} \over 32 \pi^2 m} \varepsilon_{\alpha \beta \gamma \delta}
\hk{tr}_c  F_{\alpha \beta} F_{\gamma \delta}
\nonumber \\
&& \hspace{-64pt}
 + { \hk{i} \over 192 \pi^2 m^3}
\varepsilon_{\alpha \beta \gamma \delta}
\hk{tr}_c \Big[ 2 \Big[ \nabla_\rho , \Big[ \nabla_\rho ,
F_{\alpha \beta} \Big] \Big]
F_{\gamma \delta}
+ \Big[ \nabla_\rho ,  F_{\alpha \beta} \Big]
\Big[ \nabla_\rho ,  F_{\gamma \delta} \Big]
\nonumber \\ && \hspace{32pt}
+ 4 \hk{i} F_{\rho \alpha} F_{\rho \beta} F_{\gamma \delta} \Big]
 + {\cal O}\left({1 \over m^5}\right) \, .
\eeq
{}From the Bianchi identity (\ref{eq:Bianchi})
we obtain the following relations\footnote{
In the calculation of \cite{AraMusTok98} the factor of 2
was missing in the first identity, which led to the wrong result.}
\beq
&&
\Big[ \nabla_\rho , \Big[ \nabla_\rho ,
F_{\alpha \beta} \Big] \Big]
=
-2 \hk{i} \Big[ F_{\rho \alpha} , F_{\rho \beta}  \Big]
+ \Big[ \nabla_\alpha , \Big[ \nabla_\rho , F_{\rho \beta} \Big] \Big]
- \Big[ \nabla_\beta , \Big[  \nabla_\rho , F_{\rho \alpha} \Big] \Big]
\, ,
\\ &&
\varepsilon_{\alpha \beta \gamma \delta} \hk{tr}_c
\Big[ \nabla_\rho , F_{\rho \beta} \Big]
\Big[ \nabla_\alpha, F_{\gamma \delta} \Big]
= 0 \, ,
\eeq
so that
\beq
\varepsilon_{\alpha \beta \gamma \delta} \hk{tr}_c
\Big[ \nabla_\rho , \Big[ \nabla_\rho ,
F_{\alpha \beta} \Big] \Big] F_{\gamma \delta}
&=& -2 \hk{i} \varepsilon_{\alpha \beta \gamma \delta} \hk
{tr}_c
\Big[ F_{\rho \alpha} F_{\rho \beta} F_{\gamma \delta}
- F_{\rho \beta} F_{\rho \alpha} F_{\gamma \delta} \Big]
\nonumber \\ && \hspace{-64pt}
+ \varepsilon_{\alpha \beta \gamma \delta} \hk{tr}_c
\Big[ \Big[ \nabla_\alpha , \Big[ \nabla_\rho , F_{\rho \beta} \Big] \Big]
F_{\gamma \delta}
 - \Big[ \nabla_\beta , \Big[  \nabla_\rho , F_{\rho \alpha} \Big] \Big]
F_{\gamma \delta} \Big]
\nonumber  \\ && \hspace{-80pt}
= \;
\varepsilon_{\alpha \beta \gamma \delta}
 \hk{tr}_c \Big[ - 4 \hk{i}
F_{\rho \alpha} F_{\rho \beta} F_{\gamma \delta}
+ 2 \Big[ \nabla_\alpha , \Big[ \nabla_\rho , F_{\rho \beta} \Big] \Big]
F_{\gamma \delta} \Big]
\nonumber \\ && \hspace{-80pt}
= \;
\varepsilon_{\alpha \beta \gamma \delta} \left(
 - 4 \hk{i} \hk{tr}_c
F_{\rho \alpha} F_{\rho \beta} F_{\gamma \delta}
+ 2 \partial_\alpha
\hk{tr}_c \Big[ \nabla_\rho , F_{\rho \beta} \Big]
 F_{\gamma \delta} \right)
\, .
\eeq
On the other hands it yields
\beq
\varepsilon_{\alpha \beta \gamma \delta}
\hk{tr}_c \Big[ \Big[ \nabla_\rho , \Big[ \nabla_\rho ,
F_{\alpha \beta} \Big] \Big]
F_{\gamma \delta}
+ \Big[ \nabla_\rho ,  F_{\alpha \beta} \Big]
\Big[ \nabla_\rho ,  F_{\gamma \delta} \Big] \Big]
&=& \varepsilon_{\alpha \beta \gamma \delta}
\hk{tr}_c \Big[ \nabla_\rho , \Big[ \nabla_\rho ,
F_{\alpha \beta} \Big] F_{\gamma \delta} \Big]
\nonumber \\  && \hspace{-80pt} = \;
\varepsilon_{\alpha \beta \gamma \delta}
\partial_\rho \hk{tr}_c \Big[ \nabla_\rho ,
F_{\alpha \beta} \Big] F_{\gamma \delta}
\nonumber \\ && \hspace{-80pt} = \;
\varepsilon_{\alpha \beta \gamma \delta}
\partial_\rho \tr{c}{
\Big[ \nabla_\rho , F_{\alpha \beta}  F_{\gamma \delta}\Big]
- F_{\alpha \beta} \Big[ \nabla_\rho , F_{\gamma \delta}\Big]}
\nonumber \\ && \hspace{-80pt} = \;
{1 \over 2} \varepsilon_{\alpha \beta \gamma \delta}
\partial_\rho \hk{tr}_c
\Big[ \nabla_\rho , F_{\alpha \beta}  F_{\gamma \delta}\Big]
\nonumber \\ && \hspace{-80pt} = \;
{1 \over 2} \varepsilon_{\alpha \beta \gamma \delta}
\partial^2
\hk{tr}_c F_{\alpha \beta}  F_{\gamma \delta} \, .
\eeq
So we finally end up with
\beq
\hk{tr}_{c,\gamma}\langle x | {1 \over D }
\gamma_5 | x \rangle
&=&
{ \hk{i} \over 32 \pi^2 m}
\varepsilon_{\alpha \beta \gamma \delta}
\hk{tr}_c F_{\alpha \beta} F_{\gamma \delta}
\nonumber \\ && \hspace{-64pt}
 + { \hk{i} \over 384 \pi^2 m^3}
\varepsilon_{\alpha \beta \gamma \delta}
\partial^2
\hk{tr}_c F_{\alpha \beta}  F_{\gamma \delta}
+ { \hk{i} \over 96 \pi^2 m^3}
\varepsilon_{\alpha \beta \gamma \delta}
\partial_\alpha
\hk{tr}_c \Big[ \nabla_\rho , F_{\rho \beta} \Big]
F_{\gamma \delta} + {\cal O}\left({1 \over m^5}\right)
\nonumber \\ && \hspace{-80pt} = \;
{ \hk{i} \over 16 \pi^2 m}
\hk{tr}_c F_{\alpha \beta} \tilde{F}_{\alpha \beta}
\nonumber \\ && \hspace{-64pt}
+ { \hk{i} \over 192 \pi^2 m^3}
\partial^2
\hk{tr}_c F_{\alpha \beta} \tilde{F}_{\alpha \beta}
+ { \hk{i} \over 48 \pi^2 m^3}
\partial_\alpha
\hk{tr}_c \Big[ \nabla_\rho , F_{\rho \beta} \Big]
\tilde{F}_{\alpha \beta}
+ {\cal O}\left({1 \over m^4}\right)  \, ,
\label{eq:axres}
\eeq
where we have introduced the common notation
$\tilde{F}_{\alpha \beta}=\frac{1}{2}
\varepsilon_{\alpha \beta \gamma \delta}
F_{\gamma \delta}$.

\subsubsection{The divergency of the axial vector current}
\label{axivecexp}
Instead of expanding the axial vector current
$j^5_\mu(x)=Q^\dag(x) \gamma_\mu
\gamma_5 Q(x)$ in the way outlined,
we can use that the divergence
of the axial vector current is given
by
\be
\label{eq:anomaly}
\partial_\mu j^5_\mu =
2 m Q^\dag \gamma_5 Q -  { \hk{i} \over 16 \pi^2 }
F^{a}_{\mu \nu} \tilde{F}^a_{\mu \nu} \, ,
\ee
where the first term contains the axial current and the second is the
axial anomaly term which arises due to quantum effects.
The expansion of the divergence
of the axial vector current in terms of the inverse of the heavy
quark mass is therefore reduced to the expansion
of the axial current, which we have already performed before.

Further the axial anomaly equation (\ref{eq:anomaly}) has some
general properties, which can be used to check our result
for the axial current:

First the rhs of eq.~(\ref{eq:anomaly}) vanishes in the limit
of infinite quark mass. This can be understood by the fact
that the regulator mass
cancels the physical mass in the infinite mass limit because of
the different
sign in the definition of the regulator.
Therefore we expect the order ${\cal O}(1/m)$ term in the expectation
value of the axial current multiplied by $2m$ exactly to cancel the anomaly
term. Indeed equation (\ref{eq:axres}) gives:
\beq
2 m \hk{tr}_{c,\gamma}{\langle x | {1 \over D }
\gamma_5 | x \rangle}^{{\cal O}(1/m)}
&=&
{ \hk{i} \over 8 \pi^2}
\hk{tr}_c F_{\mu \nu} \tilde{F}_{\mu \nu}\, .
\eeq

Second if one thinks of the expectation value of the axial vector current
as a local large $m$ expansion in the background of gluon fields
\be
\langle j^5_\mu(x) \rangle = \sum_{n}
{1 \over m^n} X_{\mu 5}^{(n)}(x)
\ee
then due to equation (\ref{eq:anomaly}) the expectation value of the axial
current in the large $m$ expansion is
\be
2 m \hk{tr}_{c,\gamma}{\langle x | {1 \over D }
\gamma_5 | x \rangle}
= \sum_{n}
{1 \over m^{n}} \partial_\mu X_{\mu 5}^{(n)}(x)\,.
\ee
This means that terms  appearing in the expansion of the axial current
must be of the form of a total derivative.
The order ${\cal O}(1/m^3)$ term in equation (\ref{eq:axres}) exactly
obeys this form
\beq
\label{eq:Rores}
2 m \hk{tr}_{c,\gamma}{\langle x | {1 \over D }
\gamma_5 | x \rangle}^{{\cal O}(1/m^3)}
&=&
{ \hk{i} \over 96 \pi^2 m^2 } \partial_\mu R_\mu \, ,
\\
R_\mu &=& \partial_\mu \hk{tr}_c F_{\alpha \beta} \tilde{F}_{\alpha \beta}
+ 4 \hk{tr}_c \Big[ \nabla_\alpha , F_{\alpha \nu} \Big] \tilde{F}_{\mu \nu} \,.
\label{eq:Rures}
\eeq
The term $f^{abc} F^a_{\mu \nu} \tilde{F}^b_{\nu \alpha}
F^c_{\alpha \mu}$ appearing falsely in the expansion of the axial current in
\cite{HalZhi97,AraMusTok98}
cannot be represented as a total derivative of a local
expression\footnote{A straightforward calculation
 for the instanton field shows that $\int \hk{d}^4 x
f^{abc} F^a_{\mu \nu} \tilde{F}^b_{\nu \alpha}
F^c_{\alpha \mu} \neq 0$. But for dimensional reasons this nonvanishing
contribution can be excluded from being generated by a surface term
if the instanton field is taken in the regular gauge.
Therefore the integrand cannot be a total derivative.}
 and therefore violates the general argument given above.

The expectation value for the divergence of the axial vector current
in the background of
gluon fields finally reads up to order ${\cal O}(1/m^4)$
\beq
\langle \partial_\mu j^5_\mu(x) \rangle
&=&
{ \hk{i} \over 96 \pi^2 m^2 } \partial_\mu \left(
\partial_\mu \hk{tr}_c F_{\alpha \beta} \tilde{F}_{\alpha \beta}
+ 4 \hk{tr}_c \Big[ \nabla_\alpha , F_{\alpha \nu} \Big] \tilde{F}_{\mu \nu}
\right)
+ {\cal O}\left({1 \over m^4}\right) \, . \;
\label{eq:divaxvecres}
\eeq

\subsubsection{The scalar current}
\label{scaexp}

Following the steps outlined
in the introduction to this section
the expansion of a scalar current in
series of the inverse heavy quark mass yields up
to order ${\cal O}(1/m^3)$:
\beq
\hk{tr}_{c,\gamma}\langle x | {1 \over D }
| x \rangle
&=&
\hk{tr}_{c,\gamma} \kint {1 \over k^2 + m^2}
\sum_{n=0}^{\infty} {\left( { \nabla^2 - {\sigma \over 2} F + 2 \hk{i} k\nabla
\over k^2 + m^2} \right)}^n
\left( \hk{i}\! \sla{\nabla} - \sla{k} - \hk{i} m \right)
\cdot 1 \;\;
\nonumber \\ && \hspace{-80pt} = \;
{ - \hk{i} \over {\left( 4 \pi \right)}^{\frac{d}{2}}}
{\left( { 1 \over m^2 }\right)}^{1-\frac{d}{2}} m
\Gamma\left(1-\frac{d}{2}\right) d \hk{tr}_c 1
\nonumber \\  && \hspace{-64pt}
- { \hk{i} \over 24 \pi^2 m }
\hk{tr}_c F_{\alpha \beta} F_{\alpha \beta}
\nonumber \\  && \hspace{-64pt}
+ { 1 \over 360 \pi^2 m^3}
\hk{tr}_c F_{\alpha \beta} F_{\alpha \gamma}
F_{\beta \gamma}
- { 7 \hk{i} \over 2880 \pi^2 m^3}
\partial^2
\hk{tr}_c F_{\alpha \beta} F_{\alpha \beta}
\nonumber \\  && \hspace{-64pt}
- { \hk{i} \over 720\pi^2 m^3}
\left( 11 \hk{tr}_c \Big[ \nabla_\alpha ,
\Big[ \nabla_\beta , F_{\beta \gamma} \Big] \Big]
F_{\alpha \gamma}
- \hk{tr}_c \Big[ \nabla_\alpha ,
F_{\alpha \beta} \Big]
\Big[ \nabla_\gamma , F_{\gamma \beta} \Big] \right) \, .
\label{eq:scares}
\eeq
The infinite constant term can be cancelled by substracting
the expectation value of the scalar current
for vanishing gluonic background fields.
Our result eq.~(\ref{eq:scares}) coincides with
that obtained in ref.~\cite{BroGen} if we
neglect the total derivative terms which were ignored in
ref.~\cite{BroGen}.

Actually the result eq.~(\ref{eq:scares})
with
the total derivative terms neglected
(and hence that of ref.~\cite{BroGen})
can be easily obtained from the expansion of
the determinant of the Dirac operator (\ref{eq:detall}), since
\beq
\int \hk{d}^4 x \hk{tr}_{c,\gamma}\langle x | {1 \over D }
| x \rangle
&=& { \hk{d} \over \hk{d}m}
\left( - \hk{i} \ln \left( \hk{det} D \right) \right)
\nonumber \\
&=&
{ \hk{d} \over \hk{d}m}
\left( - \hk{i} \hk{Tr} \ln D \right) \, .
\eeq
Our expansion of the scalar current (\ref{eq:scares}) is in agreement
with the result for the determinat in equation (\ref{eq:detall}).

\subsubsection{The vector current}
\label{vecexp}

The heavy quark expansion of the vector current up to order $1/m^3$
gives exactly zero
\be
\hk{tr}_{c,\gamma}\langle x | {1 \over D }
\gamma_\mu | x \rangle = 0 + {\cal O}\left({1 \over m^4}\right) \,.
\ee
This result can be easily anticipated from the fact that the vector
current is $C-$parity odd. This implies that the first operator
contributing to heavy quark mass expansion should contain at least
three gluon fields,
additionaly the vector current conservation
requires that this operator has the following structure $\nabla G^3$.
{}From counting of dimensions we see that such operator can contribute
only at $1/m^4$ order.

\subsubsection{The tensor current}
\label{tenexp}

For the color singlet tensor current we
find that the first non-vanishing order
of the expansion is ${\cal O}\left(1/m^3\right)$, yielding
\beq
\hk{tr}_{c,\gamma}\langle x | {1 \over D }
\sigma_{\mu \nu} | x \rangle
&=&
{ \hk{i} \over 24 \pi^2} {1 \over m^3} \nonumber \\
&& \times
\hk{tr}_c \Big[
F_{\alpha \beta} F_{\alpha \beta} F_{\mu \nu}
+ F_{\alpha \nu} F_{\beta \mu} F_{\alpha \beta}
- F_{\alpha \mu} F_{\beta \nu} F_{\alpha \beta} \Big]
+ {\cal O}\left({1 \over m^5}\right) \,.
\label{eq:tenres}
\eeq
We note that the rhs of the above equation vanishes in the case
of $SU(2)$ gauge group. Actually one can show that the lhs of
eq.~(\ref{eq:tenres}) is identically zero in the case of $SU(2)$
gauge group. Therefore the fact that rhs of eq.~(\ref{eq:tenres})
vanishes for $SU(2)$ gauge group is a powerful check of our
calculations.

In order to prove that lhs of eq.~(\ref{eq:tenres}) is zero
in  the case of SU(2) gauge group we use the following transformation:

$$
G=C\tau^2\, ,
$$
where $C$ is charge conjugation matrix in Dirac spinor space
and $\tau^2$ is color $SU(2)$ matrix. Under this transformation
we have:
\beq
\nonumber
G\tau^aG^{-1}&=&-\tau^{aT}\\
\nonumber
G\sigma_{\mu\nu} G^{-1}&=&-\sigma_{\mu\nu}^{T}\\
\nonumber
GD G^{-1}&=&D^{T}
\eeq
where $^T$ is the transposition operation.
The lhs.\ of eq.~(\ref{eq:tenres}) should be zero,
since
\beq
\hk{tr}_{c,\gamma}\langle x | {1 \over D }
\sigma_{\mu \nu} | x \rangle   & = &
\hk{tr}_{c,\gamma}\langle x |G {1 \over D }
\sigma_{\mu \nu} G^{-1}| x \rangle
\nonumber \\  &=&
\hk{tr}_{c,\gamma}\langle x |
\left( {1 \over D } \right)^T \left( - \sigma_{\mu \nu}^T \right)
| x \rangle
\nonumber \\  &=&
- \hk{tr}_{c,\gamma}\langle x | {1 \over D }
\sigma_{\mu \nu} | x \rangle \, .
\eeq

Nullification of the heavy quark mass expansion of the tensor current
for $SU(2)$ gauge group implies that lhs.\ of eq.(\ref{eq:tenres})
is zero if it is computed in the field of single instanton.

\subsubsection{Expansion of $\langle \bar{Q} \nabla_\mu \gamma_\nu Q \rangle$.}
\label{enemomexp}

The energy-momentum tensor of QCD can be written in Minkowski-space as
\be
T^{\mu \nu}=-g^{\mu \nu} {\cal L}_{\rm QCD} - F^{\mu \alpha} F^\nu_\alpha
+ \frac{\rm i}{2} \bar{\psi}
\partialboth^{(\mu} \gamma^{\nu)} \psi \, ,
\label{enemomten}
\ee
where $(\mu \nu)$ denotes the symmetrization of the indices.
The large $m$ expansion of the (not symmetrized) last term
in eq.~(\ref{enemomten}) yields in Euclidean space
\beq
&&\hk{tr}_{c,\gamma}\langle x | {1 \over D } \nabla_\mu \gamma_\nu | x \rangle
=
{ - 2 \, \hk{i} \over {( 4 \pi )}^{\frac{d}{2}} }
{\left( { 1 \over m^2 } \right)}^{-\frac{d}{2}}
\Gamma \left( -\frac{d}{2} \right) \delta_{\mu \nu} \hk{tr}_c 1
\nonumber \\
&&
+ { \hk{i} \over {( 4 \pi )}^{\frac{d}{2}} }
{\left( { 1 \over m^2 } \right)}^{2-\frac{d}{2}}
\Gamma \left(2- \frac{d}{2} \right)
\left( -\frac{1}{3} \delta_{\mu \nu} \hk{tr}_c F_{\alpha \beta} F_{\alpha \beta}
+ \frac{4}{3} \hk{tr}_c F_{\alpha \nu} F_{\alpha \mu} \right)
\nonumber \\
&&
+ { 1 \over 720 \pi^2 } {1 \over m^2}
\delta_{\mu \nu} \hk{tr}_c
F_{\alpha \beta} F_{\beta \gamma} F_{\gamma \alpha}
- {7 \hk{i} \over 5760  \pi^2 } { 1 \over m^2 }
\delta_{\mu \nu}
\partial^2
\hk{tr}_c F_{\alpha \beta} F_{\alpha \beta}
\nonumber \\
&&
- {\hk{i} \over 1440  \pi^2 } { 1 \over m^2 }
\delta_{\mu \nu}
\left( 11 \, \hk{tr}_c \Big[ \nabla_{\alpha} , \Big[ \nabla_{\beta} ,
F_{\beta \gamma} \Big] \Big] F_{\alpha \gamma}
-  \hk{tr}_c \Big[ \nabla_{\alpha} , F_{\alpha \beta} \Big]
\Big[ \nabla_{\gamma} , F_{\gamma \beta} \Big] \right)
\nonumber \\
&&
+ {1 \over 2880\pi^2 } { 1 \over m^2 }
\Big( -4 \hk{tr}_c F_{\alpha \nu} F_{\beta \mu} F_{\alpha \beta}
- 4 \hk{tr}_c F_{\alpha \mu} F_{\beta \nu} F_{\alpha \beta}
\nonumber \\
&& \hspace{32pt}
-30 \hk{i} \hk{tr}_c  \Big[ \nabla_{\alpha} ,
 F_{\mu \nu} \Big] \Big[ \nabla_{\beta} , F_{\alpha \beta} \Big]
\nonumber \\
&& \hspace{32pt}
+74 \hk{i} \hk{tr}_c \Big[ \nabla_{\alpha} ,
\Big[ \nabla_{\beta} , F_{\beta \nu} \Big] \Big] F_{\alpha \mu}
+14 \hk{i} \hk{tr}_c \Big[ \nabla_{\alpha} ,
\Big[ \nabla_{\beta} , F_{\beta \mu} \Big] \Big] F_{\alpha \nu}
\nonumber \\
&& \hspace{32pt}
- 26  \hk{i} \hk{tr}_c \Big[ \nabla_\nu ,
\Big[ \nabla_{\alpha} ,  F_{\alpha \beta} \Big] \Big]   F_{\beta \mu}
- 26  \hk{i} \hk{tr}_c \Big[ \nabla_\mu ,
\Big[ \nabla_{\alpha} ,  F_{\alpha \beta} \Big] \Big]   F_{\beta \nu}
\nonumber \\
&& \hspace{32pt}
+22 \hk{i} \partial^2 \hk{tr}_c
F_{\beta \nu} F_{\beta \mu}
-3 \hk{i} \partial_\mu \partial_\nu \hk{tr}_c F_{\alpha \beta} F_{\alpha \beta}
\nonumber \\
&& \hspace{32pt}
-26 \hk{i} \hk{tr}_c \Big[ \nabla_\mu ,
F_{\alpha \nu} \Big] \Big[ \nabla_{\beta}, F_{\alpha \beta} \Big]
-26 \hk{i} \hk{tr}_c \Big[ \nabla_\nu ,
F_{\alpha \mu} \Big] \Big[ \nabla_{\beta}, F_{\alpha \beta} \Big]
\nonumber \\
&& \hspace{32pt}
+ 40 \hk{i} \hk{tr}_c \Big[  \nabla_{\alpha} ,
 F_{\alpha \nu} \Big] \Big[ \nabla_{\beta} , F_{\beta \mu} \Big]
+ 4 \hk{i} \hk{tr}_c \Big[ \nabla_\nu ,
 F_{\alpha \beta} \Big] \Big[ \nabla_\mu , F_{\alpha \beta} \Big]
\Big)
+ {\cal O}\left({1 \over m^4}\right)
\label{eq:enemomres1}
\eeq
and
\be
\hk{tr}_{c,\gamma}\langle x | {1 \over D } \nabla_\mu \gamma_\nu
| x \rangle^{{\cal O}(m^0,m^{-2})}
=
\hk{tr}_{c,\gamma}\langle x| \nabla_\mu {1 \over D } \gamma_\nu
| x \rangle^{{\cal O}(m^0,m^{-2})} \, .
\label{eq:enemomres2}
\ee
The Lorentz trace of eq.~(\ref{eq:enemomres1}) can be compared with
the expansion of the scalar current in eq.~(\ref{eq:scares}) since
\be
\hk{tr}_{c,\gamma}\langle x |
{1 \over D } \nabla_\nu \gamma_\nu
| x \rangle
= - m \hk{tr}_{c,\gamma}\langle x | {1 \over D } | x \rangle \, .
\ee
For the trace of the rhs of eq.~(\ref{eq:enemomres1}) we find
\beq
 \hk{tr}_{c,\gamma}\langle x |
{1 \over D } \nabla_\nu \gamma_\nu
| x \rangle^{{\cal O}(m^0,m^{-2})}
&=&
 { \hk{i} \over 24 \pi^2 }
\hk{tr}_c F_{\alpha \beta} F_{\alpha \beta}
\nonumber \\  && \hspace{-160pt}
- { 1 \over 360 \pi^2 m^2}
\hk{tr}_c F_{\alpha \beta} F_{\alpha \gamma}
F_{\beta \gamma}
+ { 7 \hk{i} \over 2880 \pi^2 m^2}
\partial^2
\hk{tr}_c F_{\alpha \beta} F_{\alpha \beta}
\nonumber \\  && \hspace{-160pt}
+ { \hk{i} \over 720 \pi^2 m^2}
\left( 11 \hk{tr}_c \Big[ \nabla_\alpha ,
\Big[ \nabla_\beta , F_{\beta \gamma} \Big] \Big]
F_{\alpha \gamma}
- \hk{tr}_c \Big[ \nabla_\alpha ,
F_{\alpha \beta} \Big]
\Big[ \nabla_\gamma , F_{\gamma \beta} \Big] \right) \, ,
\label{eq:conenemomres}
\eeq
which exactly coincides with the rhs
of eq.~(\ref{eq:scares}) multiplied
by $(-m)$.
Eq.~(\ref{eq:enemomres2}) further
agrees with the expansion of the vector current
beeing zero up to ${\cal O}(1/m^3)$ since
\be
\hk{tr}_{c,\gamma}\langle x | \nabla_\mu {1 \over D } \gamma_\nu
| x \rangle -
\hk{tr}_{c,\gamma}\langle x | {1 \over D } \nabla_\mu \gamma_\nu
| x \rangle
= \partial_\mu \hk{tr}_{c,\gamma}\langle x | {1 \over D } \gamma_\nu
| x \rangle \, .
\ee

\section{Intrinsic heavy quarks in light hadrons}

In light hadron processes
heavy quarks may give contributions only through virtual
effects which are supressed by the mass of the heavy quarks.
Especially for the charm quark, whose mass $m_c \approx 1.4$ GeV
is not too large, virtual processes nevertheless
may give not negligible contributions.
In this section we discuss the applications of heavy quark mass
expansion obtained in the previous sections.

In the following sections we use the ``perturbative"
normalization for the gluon field
strength $G_{\mu \nu}^a=F_{\mu \nu}^a/g$ and rotate all expressions
to Minkowsky space, see the appendix.

\subsection{Intrinsic charm in $\eta$ and $\eta'$}

For the decay of the $B$-meson into $\eta'$ and $K$-mesons in
\cite{HalZhi97} a mechanism
with virtual charm quarks was suggested.
In this approach the Cabbibo favored process $b \rightarrow \bar{c}cs$
is followed by the conversion of the $\bar{c}c$ pair
directly into $\eta'$.
Its contribution to the decay amplitude is therefore
direct depending on the "intrinsic charm" component of
the $\eta'$-meson which is usually characterized by the matrix element
\be
\label{eq:deff}
\langle 0 | \bar{c} \gamma_\mu \gamma_5 c | \eta' (q) \rangle
= \hk{i} f_{\eta'}^{(c)} q_\mu \; .
\ee

Using the heavy mass expansion of the divergence of the axial vector current
(\ref{eq:divaxvecres}), the constant $f_{\eta'}^{(c)}$ can be
expressed up to the order $1/m_c^2$ by
\be
\label{eq:fanares}
f_{\eta'}^{(c)} = - {1 \over 12 m_c^2}
\langle 0 | { \alpha_s \over 4 \pi}
G^a_{\mu \nu} \tilde{G}^{\mu \nu,a} | \eta' \rangle \, .
\ee
Here we neglected the term proportional to
$\Big[ \nabla_\alpha , G_\nu^\alpha \Big]$ in (\ref{eq:divaxvecres})
which vanishes in pure Yang-Mills theory.
We now can estimate the value of the constant $f_{\eta'}^{(c)}$:
\be
\label{eq:fnumres}
f_{\eta'}^{(c)} \approx - 2 \hk{MeV} \, ,
\ee
where we have used
\be
\langle 0 | { \alpha_s \over 4 \pi}
G^a_{\mu \nu} \tilde{G}^{\mu \nu,a} | \eta' \rangle
= 0.056 \hk{GeV}^3 \, ,
\ee
obtained in \cite{DiaEid81}.

In QCD the omitted term can be related to the matrix element
\be
\label{eq:omiter}
{ \alpha_s \over 4 \pi} \langle 0 |
g \sum_{q=u,d,s} \bar{q} \gamma_\nu \tilde{G}^\nu_\mu q  | \eta' \rangle
\ee
using the equation of motion.
A rough order of magnitude estimate for the contribution of the omitted
term (\ref{eq:omiter}) to $f_{\eta'}^{(c)}$ using the results of
\cite{BalPolWei98} indicates an
deviation at the level of $0.3$ MeV to the value (\ref{eq:fnumres}).
A more careful analysis of the omitted matrix element (\ref{eq:omiter}) can be
done by using the instanton methods developed in
\cite{DiaPolWei96} which already have been applied by \cite{BalPolWei98}
to calculations of higher twist corrections to deep-inelastic scattering.

Our estimated value for $f_{\eta'}^{(c)}$ is consistent with the
phenomenological analysis in \cite{FelKro98} where the authors dervived
the bound $-65 \hk{MeV} \leq f_{\eta'}^{(c)} \leq 15 \hk{MeV}$ from the
analysis of $\gamma \eta'$ transition form
factors. From the analysis of $(\eta,\eta',\eta_c)$-mixing in \cite{FelKroSte98}
the small value $f_{\eta'}^{(c)} = -(6.3 \pm 0.6)$ MeV was derived,
taking into account off-shellness effects in the $\bar{c}c$ component
of $\eta'$ also, the value $|f_{\eta'}^{(c)}| \approx 2.4$ MeV was
found in \cite{AhmKou99}.
Further our value for $f_{\eta'}^{(c)}$ is in agreement with the
phenomenological bound $|f_{\eta'}^{(c)}| < 12$ MeV, obtained in
\cite{Pet98}, and corresponds to the result
$f_{\eta'}^{(c)} \approx -2.3$ MeV presented in \cite{Ali98}.
In Ref. \cite{Ali98} the divergence of the axial vector current
was computed using the triangle graph for the axial anomaly with
massive fermions, neglecting possible
$1/m_c^2$ contributions like
\be
f^{abc} G^a_{\mu \nu}
\tilde{G}^b_{\nu \alpha} G^c_{\alpha \mu} \;
\label{3fs}
\ee
from higher order diagrams.
Indeed our calculation shows that such "truly nonabelean"
operators do not contribute to the order $1/m_c^2$ and
our result (\ref{eq:fanares}) therefore is exactly given by
the first term of the expansion in $1/m_c^2$ of the triangle graph
\cite{PolSchTer98}.

The small value (\ref{eq:fnumres}) for $f_{\eta'}^{(c)}$ implies that the
$b \rightarrow \bar{c}cs$ mechanism does not play a major role in the
$B \rightarrow K \eta'$ decay mode.

Bigger values of $f_{\eta'}^{(c)}$ due to the operator
(\ref{3fs}) in the expansion
of the axial current up to order $1/m_c^3$ have been given
in \cite{HalZhi97}, where $f_{\eta'}^{(c)} \approx (50 - 180)$ MeV
and in \cite{AraMusTok98} with $f_{\eta'}^{(c)} \approx
- (12.3 - 18.4)$ MeV.
These results have been used by a number of authors for the analysis
of the charm content in noncharmed hadrons
(see e.g \cite{HalZhi97b,AraMusTok98b,BloShu98,ShuZhi98}), but
since the operator (\ref{3fs}) violates general properties of the axial
anomaly and it also does not appear in explicit calculations
(see section~\ref{axivecexp}), results relying on
\cite{HalZhi97,AraMusTok98} should be reconsidered.

Analogously we can immediately estimate the constant $f_{\eta}^{(c)}$
characterizing the intrinsic charm contribution to the $\eta$-meson.
Using
\be
\langle 0 | { \alpha_s \over 4 \pi}
G^a_{\mu \nu} \tilde{G}^{\mu \nu,a} | \eta \rangle
= 0.020 \hk{GeV}^3 \, ,
\ee
obtained in \cite{DiaEid81} we find
\be
f_{\eta}^{(c)} \approx - 0.7 \hk{MeV} \, .
\label{eq:fetanum}
\ee
Since in the case of the $\eta$ meson the contribution of the
omitted term (\ref{eq:omiter}) can be of the same order as
$f_{\eta}^{(c)}$ itself, the estimate (\ref{eq:fetanum})
must be considered as a poor.

\subsection{Intrinsic charm contribution to the proton spin}
Another application of our result for the heavy quark mass
expansion of the divergency of
axial vector current (\ref{eq:divaxvecres}) has been given in
\cite{PolSchTer98}.
In this paper the authors have shown that
the intrinsic charm contribution to
the first moment of the spin structure function $g_1(x,Q^2)$ of
the nucleon is small contrary to the result of
\cite{HalZhi97b,BloShu98,AraMusTok98b}.

In ref.~\cite{PolSchTer98} it was proven that the forward matrix
element of the axial current in the leading order of heavy
quark mass expansion can be computed as:
\begin{equation}
\langle N(p,{\lambda})| \bar c \gamma_\mu\gamma_5 c(0)
|N(p,S) \rangle  =
{\alpha_s\over 48 \pi m_c^2}
\langle N(p,S)| R_{\mu}(0)
|N(p,S) \rangle\, .
 \label{jc}
\end{equation}
Here the current $R_{\mu}(0)$ is given by eq.~(\ref{eq:Rures}).
Note that the first term in $R_{\mu}$ does
not contribute to the forward
matrix element because of its gradient form,
while the contribution of the
second one is rewritten, by making use of the equation
of motion, as
matrix element of the operator
\begin{eqnarray}
\langle N(p,S)| \bar c \gamma_\mu\gamma_5 c(0)
|N(p,S) \rangle&=&\frac{\alpha_s}{12 \pi m_c^2}
\langle N(p,S)|
g\, \sum_{\scriptstyle{\rm f=u,d,s}}
\bar \psi_f\gamma_\nu \tilde G_{\mu}^{~~\nu}
\psi_f | N(p,S) \rangle \
\nonumber \\
&\equiv&
\frac{\alpha_s}{12 \pi m_c^2} 2 m_N^3 S_{\mu} f^{(2)}_S,
\label{sfc}
\end{eqnarray}
The parameter $f^{(2)}_S$ was determined before in calculations
of the power corrections to the first moment
of the singlet part of $g_1$.
QCD-sum rule calculations gave
$f^{(2)}_S=0.09$
\cite{ES},
estimates using the renormalon
approach led to
$f^{(2)}_S=\pm 0.02 $
\cite{ES2} and calculations in the instanton model of the QCD vacuum
give a result very close to that of QCD sum rule \cite{BalPolWei98}.

Inserting these numbers we get finally for  the charm axial
constant of the nucleon the estimate
\begin{equation}
g_A^{(c)}=
-\frac{\alpha_s}{12\pi} f^{(2)}_S \frac{m_N^2}{m_c^2}\approx
-5 \cdot 10^{-4}
\label{gc}
\end{equation}
with probably a 100 percent  uncertainty.
Note that this contribution is of non-perturbative origin
(therefore we call it intrinsic),
so that it is sensitive to large distances, as soon
as the factorization scale is of order $m_c$.

\subsection{Intrinsic charm contribution to the nucleon tensor charge}
Using the results of section~\ref{tenexp} we can estimate the intrinsic
charm contribution to the tensor charge of the nucleon. The tensor
charge of the nucleon is defined as:
\be
\langle N(p,S)|\bar c \sigma_{\mu\nu}\gamma_5 c(0)| N(p,S)\rangle=
2 \hk{i}\, g_T^{(c)} (p_\mu S_\nu-p_\nu S_\mu)\, .
\ee
Using the result of the section~\ref{tenexp} and the identity
$$
\sigma_{\mu\nu}\gamma_5=-\frac{\rm i}{2}\varepsilon_{\mu\nu\alpha\beta}
\sigma^{\alpha\beta}\, ,
$$
we obtain for the charm contribution to the nucleon tensor charge
the following result:
\beq
&&g_T^{(c)}=
{ 1 \over 96 \pi^2} {1 \over m_c^3 m_N^2} \nonumber \\
&& \times
\varepsilon^{\lambda\rho\mu\nu} p_\lambda S_\rho
\langle N(p,S)|
\hk{tr}_c g_s^3\Big[
G_{\alpha \beta} G^{\alpha \beta} G_{\mu \nu}
+ 2\ G_{\alpha \nu} G_{\beta \mu} G^{\alpha \beta}
\Big]|N(p,S)\rangle
+ {\cal O}\left({1 \over m_c^5}\right) \,.
\label{tencharge}
\eeq
The matrix element in the rhs of above equation can be rouhgly
estimated in the instanton vacuum using method of \cite{DiaPolWei96}.
As discussed in section~\ref{tenexp} the gluonic operator in rhs of
eq.~(\ref{tencharge}) is identically zero on one instanton. Therefore
the first nontrivial contribution one can get from
instanton--anti-instanton pair. If we compare the expression for the
charm contribution to the nucleon charge with that for axial charge
we see that the charm contribution to the tensor charge is suppressed
by additional power of $m_N/m_c$ and one power of instanton packing
fraction ($\pi^2 \bar\rho^4/\bar R^4$),
however the tensor charge is enchanced by one power
of $\alpha_s(m_c)$
\footnote{
Let us note that the expansion parameter in the heavy quark mass
expansion is $g_s G/m_c$, because the non-perturbative
gluon field strength $G\sim 1/g_s$ ($cf.$  instanton field).
Therefore $g_s(\mu)$ accompanied by gluon field strength is not
counted as suppression.}.
This allows us to make a rough estimate for the
charm contribution of the tensor charge:
\beq
g_T^{(c)}\sim  \frac{m_N}{\alpha_s(m_c) \ m_c} \frac{(N_c-2)\ \pi^2\
\bar\rho^4}{N_c\ \bar R^4}\ g_A^{(c)} \sim 10^{-4}\, .
\eeq
Factors of $N_c$ are written in a way to reproduce the large $N_c$
behaviour of the matrix element and to account for the fact that
the operator in rhs of eq.~(\ref{tencharge}) is identically zero
at $N_c=2$.

\subsection{Intrinsic charm contribution to the nucleon momentum}
The charm contribution to the nucleon momentum can be defined as:
\beq
M_2^{(c)}(\mu^2)
=\int_0^1 d x_B\ x_B\
\biggl[c(x_B) + \bar c(x_B)\biggl]
=\frac{\rm i}{2(P\cdot n)^2}
\langle N(P)| \bar c \Dirac n (n\cdot \nabla)c(0)|N(P)\rangle \, ,
\eeq
where $c(x_B)$ is the charm parton distribution normalized at the
scale $\mu$, which is assumed to be
$\mu\approx m_c$. The light cone
vector $n$ is arbitrary non-collinear to nucleon momentum $P$.

Now we can use the result of eq.~(\ref{eq:enemomres1}) in order to
estimate the charm contribution to the nucleon momentum
carried by intrinsic charm quarks.

\beq
&&M_2^{(c)}(\mu)
=
\frac{\rm i}{2(P\cdot n)^2}\Biggl[
{ \hk{i}\ \alpha_s(\mu)
 \over {4 \pi } }\frac{1}
{\left(2- \frac{d}{2} \right)}\
\frac{4}{3}\
\langle N(P)| n^\mu n^\nu
\hk{tr}_c {G^{\alpha}}_{\nu} G_{\alpha \mu} |N(P)\rangle
\nonumber \\
&&
+{ 1 \over 120 \pi^2 } {{g_s^3(\mu)} \over m_c^2}
\langle N(P)|n^\mu n^\nu\
 \hk{tr}_c G_{\alpha \nu} G_{\beta \mu} G^{\alpha \beta}
|N(P)\rangle \Biggl]
+ {\cal O}\left({1 \over m_c^4}\right)\, .
\label{eq:enemomres3}
\eeq
In derivation of this expression we neglected terms which are
proportional to $[\nabla^\alpha, G_{\alpha\beta}]$ which are
suppressed by one power of $g_s(\mu)^2$.
The first term in eq.~(\ref{eq:enemomres3}) is divergent\footnote{We show
only the most singular term} and actually is related to the mixing of
quark and gluon operators. We can rewrite eq.~(\ref{eq:enemomres3}) as
follows:  \beq &&M_2^{(c)}(\mu) =\frac{4}{3}\
\frac{\alpha_s(\mu)}{4\pi} \frac{1}{\left(2- \frac{d}{2} \right)}\
M_2^{(G)}(\mu)
\nonumber \\
&+&\frac{\rm i}{2(P\cdot n)^2}\
{ 1 \over 120 \pi^2 } {{g_s^3(\mu)} \over m_c^2}
\langle N(P)|n^\mu n^\nu\
 \hk{tr}_c G_{\alpha \nu} G_{\beta \mu} G^{\alpha \beta}
|N(P)\rangle
+ {\cal O}\left({1 \over m_c^4}\right)\, .
\label{eq:enemomres4}
\eeq
Here the first term, which is proportional to
the momentum fraction carried by gluons $M_2^{(G)}(\mu)$
accounts for extrinsic charm.
Note that the coefficient in front of this term is exactly the
leading anomalous dimension $\gamma_{qG}=4/3$
which accounts for mixing quark and gluon twist-2 operators under
QCD evolution.  The intrinsic charm contribution is given by the second
term, so that we have estimates:
\beq M_2^{(c), \hk{intrinsic}}(\mu)=
\frac{\rm i}{2(P\cdot n)^2}\
 { 1 \over 120 \pi^2 } {{g_s^3(\mu)} \over m_c^2}
\langle N(P)|n^\mu n^\nu\
 \hk{tr}_c G_{\alpha \nu} G_{\beta \mu} G^{\alpha \beta}
|N(P)\rangle
+ {\cal O}\left({1 \over m_c^4}\right)\, .
\label{eq:enemomres5}
\eeq
We see that the momentum fraction carried by intrinsic charm
in the nucleon is related to the value of nucleon matrix element:
$$
\langle N(P)|n^\mu n^\nu\
 {\rm i} g_s(\mu)^3 \hk{tr}_c G_{\alpha \nu} G_{\beta \mu} G^{\alpha \beta}
|N(P)\rangle \, .
$$
One can easily see that
this matrix element in the theory of instanton vacuum
\cite{DiaPolWei96} is zero in  one-instanton approximation, the
same as matrix element
$$
\langle N(P)| n^\mu n^\nu
g_s(\mu)^2\hk{tr}_c {G^{\alpha}}_{\nu} G_{\alpha \mu} |N(P)\rangle \, .
$$
Keeping in mind that for instanton field $G\sim 1/g_s$ we can write:
\be
\frac{
\langle N(P)|n^\mu n^\nu\
 {\rm i} g_s(\mu)^3 \hk{tr}_c G_{\alpha \nu} G_{\beta \mu} G^{\alpha \beta}
|N(P)\rangle
}{
\langle N(P)| n^\mu n^\nu
g_s(\mu)^2\hk{tr}_c G^{\alpha}_{\nu} G_{\alpha \mu} |N(P)\rangle
}= \Lambda^2\, ,
\label{lambdamass}
\ee
where $\Lambda$ is parameter of the dimension of mass whose
value can be obtained using various nonperturbative methods in QCD:
lattice calculation, QCD sum rule, theory of instanton vacuum.
Generically we expect that this mass parameter is of order of
typical strong interaction scale $\Lambda \sim 1$~GeV.
Now we can rewrite eq.~(\ref{eq:enemomres5}) in terms of this
parameter and momentum fraction carried by gluons in the
nucleon at scale $\mu \approx m_c$ as:
\beq
&&M_2^{(c), \hk{intrinsic}}(\mu)=
{ \alpha_s(\mu) \over 30 \pi } {\Lambda^2 \over m_c^2}
M_2^G(\mu)
+ {\cal O}\left({1 \over m_c^4}\right)\, .
\label{eq:enemomres6}
\eeq
If we assume that $\Lambda^2=$few GeV$^2$ than we get the estimate
for the charm contribution to the nucleon momentum:
\be
M_2^{(c), \hk{intrinsic}}(\mu)= \hk{few}\times 10^{-3}\, .
\label{chislo}
\ee
We see that the heavy quark mass expansion of local currents allows
us to reduce the problem of estimate of intrinsic charm content of the
nucleon to the calculation of the ratio (\ref{lambdamass}). The latter
ratio can be computed using various methods of nonperturbative
QCD, probably the most promising would be a calculation of this ratio
in lattice QCD.

Recent analysis of refs~\cite{HSV,SMT} gives for
$M_2^{(c), \hk{intrinsic}}$ values at the level of fraction of percent
what is in agrrement with our estimate
(\ref{chislo}).

Let us note that, since we performed the heavy quark mass expansion of
heavy quark part of
energy momentum tensor not neglecting total derivatives, one can compute
also its non-forward nucleon matrix element. From the non-forward
matrix element of energy momentum one can obtain the total angular
momentum carried by intrinsic heavy quarks in the nucleon using
Ji's sum rules \cite{Ji}. The corresponding estimates we shall report
elsewhere.

\section{Conclusions}

In this paper we have computed the heavy quark mass expansion
of various local heavy quark currents.
The details of the technique are illustrated on the example
of heavy quark mass expansion of the pseudoscalar density
$\bar Q \gamma_5 Q$. This operator plays an important role
in problems related to intrinsic charm contribution to
the proton
spin and to intrinsic charm content of $\eta,\eta'$ mesons.

We corrected the mistakes in refs.~\cite{HalZhi97,AraMusTok98}
for heavy quark mass expansion of the operator $\bar Q \gamma_5 Q$ .
In these papers large intrinsic charm contribution to
the proton spin and to intrinsic charm content of $\eta,\eta'$ mesons
was obtained due to contribution of the operator
$f^{abc} G^a_{\mu \nu}
\tilde{G}^b_{\nu \alpha} G^c_{\alpha \mu} \;$
which appeared in heavy quark mass expansion
of the operator $\bar Q \gamma_5 Q$ presented
in refs.~\cite{HalZhi97,AraMusTok98}. We showed that coefficient
in front of this operator is identically zero (the result which
actually follows from general properties of the axial anomaly
\cite{FraPobPolGoe99}), so that the physical effects
based on presence of the above operator
discussed in refs.~\cite{HalZhi97,AraMusTok98,
HalZhi97b,AraMusTok98b,BloShu98,ShuZhi98} are absent.

For the first time
we presented the full results\footnote{Not neglecting
total derivatives and terms proportional
to $[\nabla^\mu, G_{\mu\nu}]$}
for heavy quark mass expansion of
the operators $\bar Q Q$ (to the order $1/m^3$), $\bar Q \gamma_5 Q$
(to the order $1/m^3$),
$\partial^\mu \bar Q\gamma_\mu \gamma_5 Q$ (to the order $1/m^2$),
$\bar Q \gamma_\mu Q$ (to the order $1/m^3$),
$\bar Q\sigma_{\mu\nu} Q$ (to the order $1/m^3$), and
$\bar Q \gamma_\mu \nabla_\nu Q$ (to the order $1/m^2$).

The results obtained for heavy quark mass expansion allowed us to
estimate the intrinsic charm content of $\eta', \eta$ mesons as well
the charm contribution to the proton spin, nucleon tensor charge
and to the fraction of nucleon momentum carried by intrinsic charm.
In the case of charm content of $\eta',\eta$ mesons
and intrinsic charm contributions to the proton spin we reduce the
calculations of these quantities to matrix elements which are already
known either phenomenologically or were computed previously.
In other cases, like intrinsic charm contribution to the nucleon
tensor charge and to energy momentum tensor, the problem is reduced
to matrix elements of gluon operators which can be estimated
using various nonperturbative methods in QCD:
lattice calculation, QCD sum rule, theory of instanton vacuum.

We made here rough order of magnitude estimate of matrix elements
of gluon operators appearing in heavy quark mass expansion of
tensor current and of energy momentum tensor using instanton model of
QCD vacuum. More quantitative estimates will be given elsewhere.
The predictions for intrinsic charm contribution to various
observables are summarized in Table~1.  \begin{table}[h] \begin{center}
\begin{tabular}{|c|c|}
\hline
quantity & our estimate \\
\hline
\hline
$f_{\eta'}^{(c)} $ & -2 MeV\\
\hline
$f_{\eta}^{(c)}$  & -0.7 MeV\\
\hline
$g_{A}^{(c)}$  &  -$5\cdot 10^{-4}$\\
\hline
$g_{T}^{(c)}$  &  $\sim 10^{-4}$\\
\hline
$M_2^{ (c),\hk{intrinsic}}$  &  $\sim 10^{-3}$\\
\hline
\end{tabular}
\caption{
{\it
Results for intrinsic charm contribution to various observables. }
}
\end{center}
\end{table}

\noindent
{\bf Acknowledgments:}

\noindent
We gratefully acknowledge useful discussions with
F.~Araki, P.V.~Pobylitsa, A.~Sch\"afer, E.~Shuryak and O.V.~Teryaev.
The work has been supported in parts by DFG and BMFB.

\section{Appendix}
\label{Euclidization}

For the euclidization we use the follwing conventions:

\be
\begin{array}{rclrclcrcl}
\hk{i} x^{0}_{\rm M} & = & x_{4,\hk{E}} \, , & x^k_{\rm M} & = & x_{k,{\rm E}}
& \rightarrow \hk{d}^4 x_{\rm M} & = & - \hk{i} \hk{d}^4 x_{\rm E} \, ,
\\
\partial^0_{\rm M} & = & \hk{i} \partial_{4,{\rm E}} \, ,
& \partial^k_{\rm M} & = & - \partial_{k,{\rm E}} \, , &&& \\
A^0_{\rm M} & = & \hk{i} A_{4,{\rm E}} \, ,
& A^k_{\rm M} & = & - A_{k,{\rm E}} \, . &&&
\end{array}
\ee
The covariant derivative therefore reads in Minkowski and in
Euclidean space-time:
\beq
\nabla^\mu_{\rm M} &=& \left(\partial^\mu - \hk{i} A^\mu(x) \right)_{\rm M} \, ,
\\
\nabla_{\mu,{\rm E}}&=&\left(\partial_\mu - \hk{i} A_\mu(x) \right)_{\rm E} \, .
\eeq
The field strength, defined as
\be
F^a_{\mu \nu}=\partial_\mu A_\nu - \partial_\nu A_\mu
+ f^{abc} A^b_\mu A^c_\nu = \hk{i} \Big[ \nabla_\mu, \nabla_\nu \Big]
\ee
transforms as
\be
F_{ij,{\rm M}}=F_{ij,{\rm E}} \, , \; \; \; F_{0j,{\rm M}}=\hk{i}
F_{4j,{\rm E}} \, .
\ee
For the Dirac matrices we choose the conventions:
\be
\gamma^0_{\rm M}=\gamma_{4,{\rm E}} \, ,
 \;\;\; \gamma^k_{\rm M}=\hk{i} \gamma_{k,{\rm E}}\, ,
\ee
and $\gamma_5$ is defined within this paper as:
\be
\gamma_{5,{\rm M}}=\gamma^5_{\rm M}
=-\hk{i}\left(\gamma^0 \gamma^1 \gamma^2 \gamma^3
\right)_{\rm M}=\left(\gamma_1 \gamma_2 \gamma_3 \gamma_4\right)_{\rm E}
=\gamma_{5,{\rm E}} \, .
\ee
With
\be
\varepsilon^{0123}_{\rm M}=-\varepsilon_{0123,{\rm M}}
=+1=\varepsilon_{1234,{\rm E}} \, ,
\ee
it yields
\be
\hk{tr}_\gamma \Big[ \gamma_5 \gamma_\alpha \gamma_\beta \gamma_\gamma
\gamma_\delta \Big]_{\rm E}
= 4 \varepsilon_{\alpha \beta \gamma \delta,{\rm E}} \, .
\ee
The fermionic fields transform as
\be
\psi_{\rm M}=\psi_{\rm E} \, , \;\;\; \bar{\psi}_{\rm M}
 = - \hk{i}\psi_{\rm E}^\dag \, ,
\ee
so the Dirac operator in Euclidean space-time reads:
\be
D=\hk{i}\! \sla{\nabla} + \hk{i} m \, .
\ee
In section \ref{secdet} we have used the following transformation properties
for the effective action and the appearing operators:
\beq
S_{\rm eff,M}&=&\hk{i} S_{\rm eff,E} \, , \\
\left( F_{\alpha \beta} F^{\alpha \beta} \right)_{\rm M} &=&
\left( F_{\alpha \beta} F_{\alpha \beta} \right)_{\rm E} \, , \;\;\;
\left( F_{\alpha \beta} {F^\beta}_{\gamma} F^{\gamma \alpha} \right)_{\rm M} = -
\left( F_{\alpha \beta} F_{\beta \gamma} F_{\gamma \alpha} \right)_{\rm E} \, .
\eeq

\end{document}